\documentstyle[aaspp]{article}
\begin{document}

\title{SELF-COLLIMATION AND MAGNETIC FIELD GENERATION OF ASTROPHYSICAL JETS}

\author{Mitsuru Honda}
\affil{\it Center for Promotion of Computational Science and Engineering,\\
Japan Atomic Energy Research Institute, Kyoto 619-0215, Japan}

\author{Yasuko S. Honda}
\affil{\it Department of Physics, School of Science, Kwansei Gakuin University,\\
Sanda, Hyogo 669-1337, Japan}

\begin{abstract}
     A novel model for collimation and transport of
electron-positron-ion jets is presented.
Analytical results show that the filamentary structures can be sustained by
self-induced toroidal magnetic fields permeating through the filaments,
whose widths significantly expand in the pair-dominant regimes.
The magnetic field strength reflects a characteristic of
equipartition of excess kinetic energy of the jets.
It is also shown that growth of the hoselike instability is strongly suppressed.
Essential features derived from this model are consistent with
recent results observed by using the very long baseline telescopes.
\end{abstract}


\section{INTRODUCTION}
     Now, there still exist some challenging problems unsolved on astrophysical jets:
these include the acceleration (Blandford \& Payne 1982), collimation,
and stability (Benford 1978; Appl \& Camenzind 1992).
For the past decades, substantial studies have been devoted to investigating
the directional mass acceleration based on the magnetohydrodynamics
(MHD; Shibata \& Uchida 1986, 1990; Cao \& Spruit 1994; Bell 1994;
Matsumoto et al. 1996; Kudoh \& Shibata 1997a, 1997b)
and radiative (Tajima \& Fukue 1996, 1998; Fukue, Tojyo, \& Hirai 2001)
mechanism around the central engine of the accretion disk.
Koide, Nishikawa, \& Mutel (1996) have performed the numerical simulation of
a relativistic MHD (RMHD) jet injected into a poloidal magnetic field.
The three-dimensional RMHD simulations are in progress worldwide
(Nishikawa et al. 1997; Aloy et al. 1999).
In addition, Carilli \& Barthel (1996) have proposed the ballistic, pressure-matched jets
whose internal pressure was assumed to be comparable to
or even less than the external one.

     On the other hand, one can find observational evidence that
the energy density of jets is larger than that of the cocoon or
intergalactic medium (e.g.,~4C~32.69; Pottash \& Wardle 1980).
Recent rotation measures using very long baseline interferometry (VLBI)
also reveal that the direction of magnetic field vectors is likely to be
transverse to the jet axis (e.g.,~1803+784; Gabuzda 1999).
In some cases, the transverse fields are so smooth that one can barely
discriminate the knots, and
it seems difficult to wholly understand such polarization properties
by invoking a trail of oblique compressional shocks.
A puzzling question is, therefore, how to demonstrate the orientation of
magnetic fields as well as the related {\it self}-collimation mechanism,
which must sustain the large-scale structure as observed.
Nevertheless, to the best of our knowledge,
there is no publication that proposes such a coherent scenario
in the context of astrophysical jets, especially
active galactic nucleus (AGN) jets,
which extend up to megaparsec scales (Mpc~$\sim 3 \times 10^{24}~{\rm cm}$)
with narrow opening angles $(\angle \phi~\lesssim~1 ^\circ)$ in some objects
(e.g.,~NGC~6251: Bridle \& Perley 1984; Cyg~A: Carilli et al. 1998;
M87,~PKS~0637-752: Hirabayashi 2001).

    We address in this Letter that the toroidal (transverse) magnetic field
induced by electron-positron flow is a favored candidate for the collimation force
of the AGN jets.
We find that the screening effects of electron-positron-ion plasmas
play a significant role in the self-organization of the filament envelopes,
whose radii exceedingly stretch in the positron-rich regimes.
The screening by the pairs and the diffuse envelope
prevent jets from the snakelike distortion.
A scaling law of the maximum possible magnetic fields is displayed
as a matter of convenience.
We expect that the present model is also applicable to Galactic jets,
except for neutral flows from protostars, and so on (Tajima \& Shibata 1997).

     The key issue discussed here is the recurrence of
the "plasma universe" model proposed by Alfv\'en (1981),
which is relevant to cosmic ray generation and transport
involving pinches (Trubnikov 1991).
So far, the present plasma configuration itself has been
considered to be notoriously unstable,
but we attempt to rewrite that scenario.
The new points, going beyond the pioneering work by Benford (1978),
are mainly
(1) the pairs are introduced,
(2) the scaling of jet radius and field strength is presented, and
(3) detailed kinetic theory is applied to the instability analysis.
We also mention that the underlying physics is ubiquitous.
Regarding recent laboratory experiments, the high power fusion lasers up to
a petawatt ($10^{15}~{\rm W}$)
reproduce the collimated relativistic electron jets (Key et al. 1998),
several tens MeV ions (Snavely et al. 2000),
and even the Bethe-Heitler pair productions triggered by
the bremsstrahlung $\gamma$-ray photons (Cowan et al. 1999).
These experiments are now opening door to the
high-field laboratory astrophysics.

    As ordinarily expected, in fully a relativistic regime of
$T>10^{10}~{\rm K}$, ejecta could consist of electrons, positrons,
and a small portion of ions, i.e.,
$n_{e-}{\gtrsim}n_{e+}{\gg}n_i$, where $n_{e-}$, $n_{e+}$, and $n_i$ are the
number densities of electrons, positrons, and ions, respectively.
Recently, the {\it ASCA} satellite has also detected
X-ray emissions of various ion species,
such as Fe, Ni, Mg, Si, S, Ne, and Ar, from the SS~433 jets (Kotani et al. 1996).
In optically thick regions, the electron-positron photoplasmas may establish
the quasi-Wien equilibria (Iwamoto \& Takahara 2002).
Anyhow, around central engines,
fast electron-positron flows relative to ion motions can be easily organized
owing to the difference of their inertia.
Namely, the electron-positron clouds are {\it primarily} accelerated by, e.g.,
the Lorentz force:
$|\mp e(v_{e \mp}/c) \times B|/ m_{e\mp}
\gg |\langle Z^*\rangle e(v_{ion}/c) \times B|/ (\langle A \rangle m_p)$,
where $m_p/m_{e\mp}=1836$, and $\langle Z^*\rangle$ and $\langle A \rangle$ are
the averages of charge state and mass number of the multispecies ions, respectively.

     The most important point is that the stream along the jet axis
$z$ with the excess electrons
$n_{e-}^{*}=n_{e-}-n_{e+}=\langle Z^* \rangle n_i$
generates the toroidal "self"-magnetic field $B_{\theta}^{s}$,
which participates in self-pinching the electron-positron gas
and in assembling the ions radially inwards
on the hydrodynamical timescale
(Honda, Meyer-ter-Vehn, \& Pukhov 2000a, 2000b).
Since the plasma holds a quite high conductivity, a closed current system
including a return current inside and/or outside the jets is self-organized
without delay.
It is noted is that the toroidal magnetic fields act as
a defocusing force for the
forward-running beam positrons and backward return electrons,
whereas such deflections tend to be immediately restored by
the electrostatic fields due to microscopic charge separations,
to avoid unphysical charge-up.
Therefore, one can treat the electron-positron flow as
the negatively charged fluid, which partially compensates for the
positive ion background, as shown below.

\section{THE ELECTRON-POSITRON FLUID EQUATIONS}
     We start with a relativistic electron-positron beam-plasma equilibrium,
invoking that the nonrelativistic equation of state of ideal gas
$p=nT$ is valid for the relativistic bulk motion (Rindler 1982).
Taking the return components into account,
radial force balance on the electron-positron fluid elements
can be expressed as ${\partial}p_{j\mp}/{\partial}r=q_{\mp}n_{j\mp}(E_r+v_{j\mp}{\times}B_{\theta}^{s}/c)$,
where $j=b,r$ indicate the beam and return components, respectively.
For electrons and positrons, we set $q_{-}=-|e|$ and $q_{+}=|e|$, respectively.
The photon pressure is omitted, and
the optically thin jet is sufficiently distant from the central engine.
All physical quantities are in the ion rest (jet) frame throughout this Letter;
this frame is chosen even in the case of
$n_{e-}\approx n_{e+}>2 \times 10^3 \langle A \rangle n_{i}$.
The force balance equations can be then cast to

\begin{equation}
{\partial\over{\partial r} }
\left(
\begin{array}{cc}
        f_{b}n_{e-}(r)T_{b-} \\
        (1-f_{b})n_{e-}(r)T_{r-} \\
        f_{b}n_{e+}(r)T_{b+} \\
        (1-f_{b})n_{e+}(r)T_{r+}\\
      \end{array}
\right)=
\left(
\begin{array}{cc}
        -ef_{b}n_{e-}(r) & e\beta_{b-}f_{b}n_{e-}(r) \\
        -e(1-f_{b})n_{e-}(r) & e\beta_{r-}(1-f_{b})n_{e-}(r) \\
        ef_{b}n_{e+}(r) & -e\beta_{b+}f_{b}n_{e+}(r) \\
        e(1-f_{b})n_{e+}(r) & -e\beta_{r+}(1-f_{b})n_{e+}(r) \\
      \end{array}
\right) \;
\left(
\begin{array}{cc}
        E_r(r) \\
        B_{\theta}^{s}(r) \\
      \end{array}
\right),
\end{equation}

\begin{equation}
E_r(r)={4\pi\over r} \left[
\sum_{j,\mp}q_{\mp}\int_{0}^{r}dr'r'n_{j\mp}(r')
+\langle Z^{*} \rangle e\int_{0}^{r}dr'r'n_i(r')\right],
\end{equation}

\begin{equation}
B_{\theta}^{s}(r)={4\pi\over r}
\sum_{j,\mp}q_{\mp}\int_{0}^{r}dr'r'\beta_{j\mp}n_{j\mp}(r'),
\end{equation}

\noindent
where $n_{b\mp}(r)=f_{b}n_{e\mp}(r)$ and $n_{r\mp}(r)=(1-f_{b})n_{e\mp}(r)$
reflect the portions of the forward beam and backward return flows, respectively.
Furthermore, we introduce the ratio of positron/electron densities,
$0\leq f_{p}\equiv n_{e+}/n_{e-}={\rm const}<1$
and the fractional charge neutrality of ion/electron charge densities,
$0<f_{c}\equiv \langle Z^{*} \rangle n_{i}/(n_{e-}-n_{e+})={\rm const}~{\leq}~1$.
Assuming $\beta_{j\mp}(r)\equiv v_{j\mp}(r)/c={\bar\beta}_{j}$ and
$T_{j\mp}(r)={\bar T}_{j}$ (isothermally) without loss of generality,
we obtain,

$$
[f_{b}{\bar T}_{b}+(1-f_{b}){\bar T}_{r}](1+f_{p})\partial n_{e-}(r)/\partial r
=(4\pi e^2/r) \times
$$
\begin{equation}
\{(1-f_{c})- [f _{b}{\bar\beta}_{b}+(1-f_{b}){\bar\beta}_{r} ]^2 \}
(1-f_{p})^2 n_{e-}(r)\int_{0}^{r}dr'r'n_{e-}(r'),
\end{equation}

\noindent
which can be applied to the cold plasma (electron-ion) jets for $f_{p}\ll 1$,
electron-positron-ion jets for $f_{p}\lesssim 1$,
and electron-positron jets for $f_{p}\simeq 1$.

\section{ANALYTICAL SOLUTIONS: JET COLLIMATION AND TOROIDAL MAGNETIC FIELD}
     The master equation~(4) can be self-consistently solved
for the total electron density $n_{e-}(r)$.
In the case of $[f _{b}{\bar\beta}_{b}+(1-f_{b}){\bar\beta}_{r}]^2>1-f_{c}$,
implying that the confinement force of the self-generated magnetic field
exceeds the repulsive force due to the space charge,
we find the solution in the form of the newly modified Bennett equilibrium:
$n_{e-}(r)=n_{e-}(r=0)/[1+(r/r_J)^2]^2$.
Here $r_J$ characterizes the radius of a filament of jet, which is given by

\begin{equation}
r_J\equiv \sqrt{ {2\pi (1+f_{p}){\bar \lambda}_D^2}\over
{ (1-f_{p})^2 \{ [f _{b}{\bar\beta}_{b}-(1-f_{b})|{\bar\beta}_{r}|]^2-(1-f_{c})} \} },
\end{equation}

\noindent
where ${\bar \lambda}_D\equiv \{ [ f_{b}{\bar T}_{b}+(1-f_{b}){\bar T}_{r} ] /{4\pi {\bar n}_{e-}}e^2  \}^{1/2}$
stands for the effective Debye sheath of electrons.
For $f_{c}\simeq 1$ and $f_p\simeq 1$,
the effective diameter of the charge-neutralized filament is
approximated by $d_J=2r_J\simeq 
\delta \bar{\beta}^{-1}\delta f_{p}^{-1} (4{\bar T}/{\bar n}_{e-}e^2)^{1/2}$.
This scales as

\begin{equation}
d_J\simeq 1.5\times 10^{12}
{{10^{-3}}\over \delta f_{p}}
{10^{-1}\over \delta \bar{\beta}}
\left ( {{\bar T}\over 10^{10}{\rm K}} \right )^{1/2}
\left ( {10^{-3} {\rm cm}^{-3}\over {\bar n}_{e-}} \right )^{1/2} ~{\rm cm},
\end{equation}

\noindent
where
$\delta f_{p}^{-1} \equiv (1-f_{p})^{-1}$,
$\delta \bar{\beta}^{-1} \equiv [f _{b}{\bar\beta}_{b}-(1-f_{b})|{\bar\beta}_{r}|]^{-1}$,
and ${\bar T}\equiv f_{b}{\bar T}_{b}+(1-f_{b}){\bar T}_{r}$
are the pair production rate, the current-neutral rate,
and the effective thermal spread, respectively.
When the return currents flow outside the filaments, i.e.,
their envelope and/or ambient medium,
the current-neutral rate reduces to
$\delta \bar{\beta}^{-1} \simeq {\bar\beta}_{b}^{-1}$.

     In Figure~1 for $f_c=1$, we show that the diameter of a filament swells
as the pair production rate and return current increase
and as the average density of the filament decreases.
Physically, the increase of positron density leads to the seeming decrease of
electron density, because only a small portion of electrons can take part in
screening ions effectively, viz.,
${\delta f}_p n_{e-}=\langle Z^{*} \rangle n_i/f_c \approx \langle Z^{*} \rangle n_i$,
and this results in significantly stretching the sheath radius.
Moreover, the return current inside the filaments
enforces to screen the magnetic fields,
so that the filaments with the lower energy density of the magnetic fields
prefer to be spread, to arrange the required magnetic confinement force
against the thermal expansion and electrostatic repulsion.

     According to self-consistent analysis of particle orbits,
the larger radius does not link with the larger current capacity (Honda 2000).
The {\it net} current inside the filament sheaths will be strongly limited because of the orbital migration of electrons embedded in self-generated magnetic fields.
That is, $i_J\simeq 4e\delta \bar{\beta}c\delta f_{p}{\bar n}_{e-}r_J^2
\lesssim 1.65{\bar\beta}_{b}{\bar\Gamma}_{b}m_ec^3/e
=28.2{\bar\beta}_{b}{\bar\Gamma}_{b}~{\rm kA} \equiv i_0$,
where ${\bar\Gamma}_{b}=(1-{\bar\beta}_{b}^2)^{-1/2}$.
This allows the parameter region of 
$1\geq \delta \bar{\beta}\delta f_{p}\gtrsim 2.4{\bar T}/{\bar\Gamma}_{b}m_ec^2$.
We conjecture that ejecta with the huge current $I_J>i_0$ (Appl \& Camenzind 1992)
necessarily split into many filaments, each carrying
one "unit" current of $\sim i_0$,
so that the number of the filaments is of the order of magnitude of $N_f\sim I_J/i_0$
(Honda et al. 2000b).
Note that the forward currents must be almost compensated with
the backward return currents.
The morphology seems to be consistent with filamentary structure
including a backward flow discovered by
the {\it Highly Advanced Laboratory for Communications and Astronomy}
VLBI Space Observatory Programme survey
(e.g.,~NGC~1275/3C~84; Asada et al. 2000), still one needs further study.

     Substituting the diffuse density profile into equation~(3), we obtain
the radial profile of magnetic field
$|B_{\theta}^{s}(r)|=2\pi e\delta \bar{\beta}\delta f_{p}n_{e-}(r=0)r/[1+(r/r_J)^2]$.
At $r=r_J$, the field takes the maximum value, to give for $f_c\simeq 1$

\begin{equation}
B_{\theta,{\rm max}}^s \simeq 0.33
\left ( 1+f _{p} \right )^{1/2}
\left ( {{\bar n}_{e-}\over 10^4 {\rm cm}^{-3}} \right )^{1/2}
\left ( {{\bar T}\over 10^{10} {\rm K}} \right )^{1/2} 
~{\rm G}.
\end{equation}

\noindent
Concerning the slowly decaying property of $|B_{\theta}^{s}(r)| \propto r^{-1}$,
the actual diameter of a jet (i.e., a bundle of the numerous filaments)
can be envisaged as
$D_J \sim 4 (N_f/\pi)^{1/2}
({B_{\theta,{\rm max}}^s}^2/8 \pi \epsilon_a)^{1/2} d_J \gg d_J$,
where $\epsilon_a$ is the energy density of the ambient medium.
In contrast with the diameter $D_J$, the maximum field
estimated in equation~(7) does not depend on $\delta \bar{\beta}$ and $N_f$.
It is instructive to notice the relation of
${\bar n}_{e-} {(\bar\Gamma}_b-1) m_e c^2 \gg {B_{\theta,{\rm max}}^s}^2/8\pi
\simeq (2/\pi){\bar n}_{e-} \bar T$,
namely, $\beta \equiv 8\pi p_{\rm th}/B^2\sim {\rm O}(1)$ (Tajima \& Shibata 1997),
consistent with equipartition of excess kinetic energy of the jet
which serves as a free energy source.

     In Figure~2, for $f_{c}=1$, we plot the range of possible magnetic fields
for a given temperature and density of jets.
The maximum by equation~(7) may be a theoretical restriction
to the field strength of the astrophysical jets,
and close to the value required from the synchrotron self-Compton (SSC) model,
which explains a double-humped appearance
on spectral energy distributions (e.g.,~Mrk~501; Kataoka et al. 1999).
On the other hand, from the spectral fitting by
the modified synchrotron proton blazar (SPB) model,
we expect a somewhat larger value
(M\"ucke \& Protheroe 2001),
which is required from the strong synchrotron radiation,
relating to stochastic acceleration of ultra-high energy cosmic rays.
Quasi-perpendicular shocks, whose configurations are favorably set up
in accordance with the present mechanism,
can probably accelerate particles up to
$10-100~{\rm EeV}$ ($10^{19}-10^{20}~{\rm eV}$).
The parameter regions of other AGN jets (e.g.,~M87, 3C~273, Cyg~A),
which are still being argued at the moment, are also shown in Figure~2.
More detailed and comprehensive observations are required
in order to clarify the detailed activities in the tips and
main bodies of jets (see e.g. Zavala \& Taylor 2002).\\

\section{DISCUSSION AND CONCLUSIONS}
     Finally, we remark that the present model is tolerant for
the synchrotron cooling and the beam-plasma--type instability,
which could be the major competitive processes.

1. Let us suppose that 
beam electrons with a terminal energy of $0.1-1$~GeV launched
from a central engine lose their energies,
by the synchrotron radiation due to their own magnetic fields,
to $1-10$~MeV, comparable to the transverse thermal spread.
In this case, the synchrotron cooling time is of the order of
$\tau_{\rm syn}\sim (10^2-10^4)(0.1~{\rm gauss}/ \bar{B})^2$~yr,
which corresponds to the propagation distance up to
$\sim 10~$kpc for ${\bar n}_{e\mp}\sim 10^4~{\rm cm}^{-3}$
and about a megaparsec for ${\bar n}_{e\mp}\lesssim 10^2~{\rm cm}^{-3}$.
In addition, reacceleration processes of the electrons, if they work,
further lengthen the propagation distance.

2. Mutual coupling between beam and return currents
as well as dissipation processes can cause various beam-plasma instabilities.
Below we briefly explain the possible mechanism of the growth rate reduction
in terms of the resistive hose instability as an example.
For the square radial profile, single frequency
seen by a beam particle can be resonant with the surface perturbation
when $\Omega \equiv \omega-kc \bar{\beta}_b \simeq
(\bar{\beta}_b \delta \bar{\beta} \delta f_p / {\bar\Gamma}_b)^{1/2}
\bar{\omega}_{pe}$,
where $\bar{\omega}_{pe} \equiv (4 \pi \bar{n}_{e-} e^2 / m_e)^{1/2}$
denotes the mean plasma frequency.
However, for the diffuse profile presented above, the growth rate is bounded,
and there is no such single frequency $\Omega$
for which the entire beam is resonant with the wave (Uhm \& Lampe 1980).
According to the self-consistent Vlasov-Maxwell analysis,
the "off-resonant" dispersion yields the most dominant wavenumber
in the complex form of
$k^*=i {\rm Im}(k_i) \pm {\rm Re}(k_r)$.
We finally get the growth distance of
$L_i \equiv 2 \pi /{\rm Im}(k_i)
\simeq \pi c {\tau}_d {\bar\beta}_b \delta \bar{\beta}
/[0.7 f_b{\bar\beta}_b + (1-f_b)|{\bar\beta}_r|]$,
where ${\tau}_d \simeq \pi r_J^2 \sigma/c^2$
stands for the magnetic diffusion time and
$\sigma$ is the electrical conductivity,
and the oscillation wavelength of 
$ \lambda_r \equiv 2 \pi /{\rm Re}(k_r)
\simeq (15 c/\bar{\omega}_{pe})
[ {\bar\Gamma}_b \bar{\beta}_b / (\delta \bar{\beta} \delta f_p) ]^{1/2}$.
Note the relation of $L_i \gg \lambda_r$,
in contrast to $L_i \sim \lambda_r$ for the resonant case.
When assuming the relativistic Spitzer conductivity
$\sigma \sim 10^{12} (T/10^{10} K)~{\rm s^{-1}}$ for $T \gg m_e c^2$
(Braams \& Karney 1989), the distance $L_i$ can be expressed as

\begin{equation}
L_i \sim 10^{22}
\left ( {10^{-3}\over \delta f_p} \right )^2
{10^{-1}\over f_b}
{10^{-3}\over \delta \bar{\beta}}
\left ( {{\bar T}\over 10^{11} {\rm K}} \right )^2
{10^{-3} {\rm cm}^{-3}\over \bar{n}_{e-}}~{\rm cm},
\end{equation}

\noindent
for $\bar{\beta}_b \simeq 1$ and $\delta f_p \ll 1$.
The pairs expand the radius $r_J$,
to significantly increase the phase lag ${\tau}_d$.
Evidently, the screening effects by the pairs lower the growth rate.
We also note that if there exist poloidal (longitudinal) magnetic fields
superposed on the transverse fields (Gabuzda \& G{\'o}mez 2001),
the shear structure of the helical fields could be well established (Miyamoto 1989).
They will play significant roles in stabilizing and guiding the jets (Davidson 1990)
and be able to further stretch the distance $L_i$.

     In conclusion, we have newly developed a generic model
for collimation and transport of relativistic electron-positron-ion jets.
A jet as a bundle of many filaments can be sustained owing to toroidal magnetic fields
self-generated by the negatively charged stream.
The magnetic field pressure balances with the gas pressure of the jet.
The pair-screening effects and return currents expand the filament widths,
and such expansion and the diffuse envelope of the filaments
lead to strong suppression of the instabilities.
In order to fully demonstrate observational results,
three-dimensional relativistic Vlasov-Maxwell or
electromagnetic particle-in-cell simulations
with the larger spatiotemporal scale should be promoted in future.

     We acknowledge useful discussions with M. Kusunose, Y. Sentoku, and A. Mizuta.
M.H. thanks the RIST and APR-JAERI for their hospitalities.
This work was supported in part by the Grants-in-Aid of ITBL-Japan.

\clearpage

\begin{figure}
\plottwo{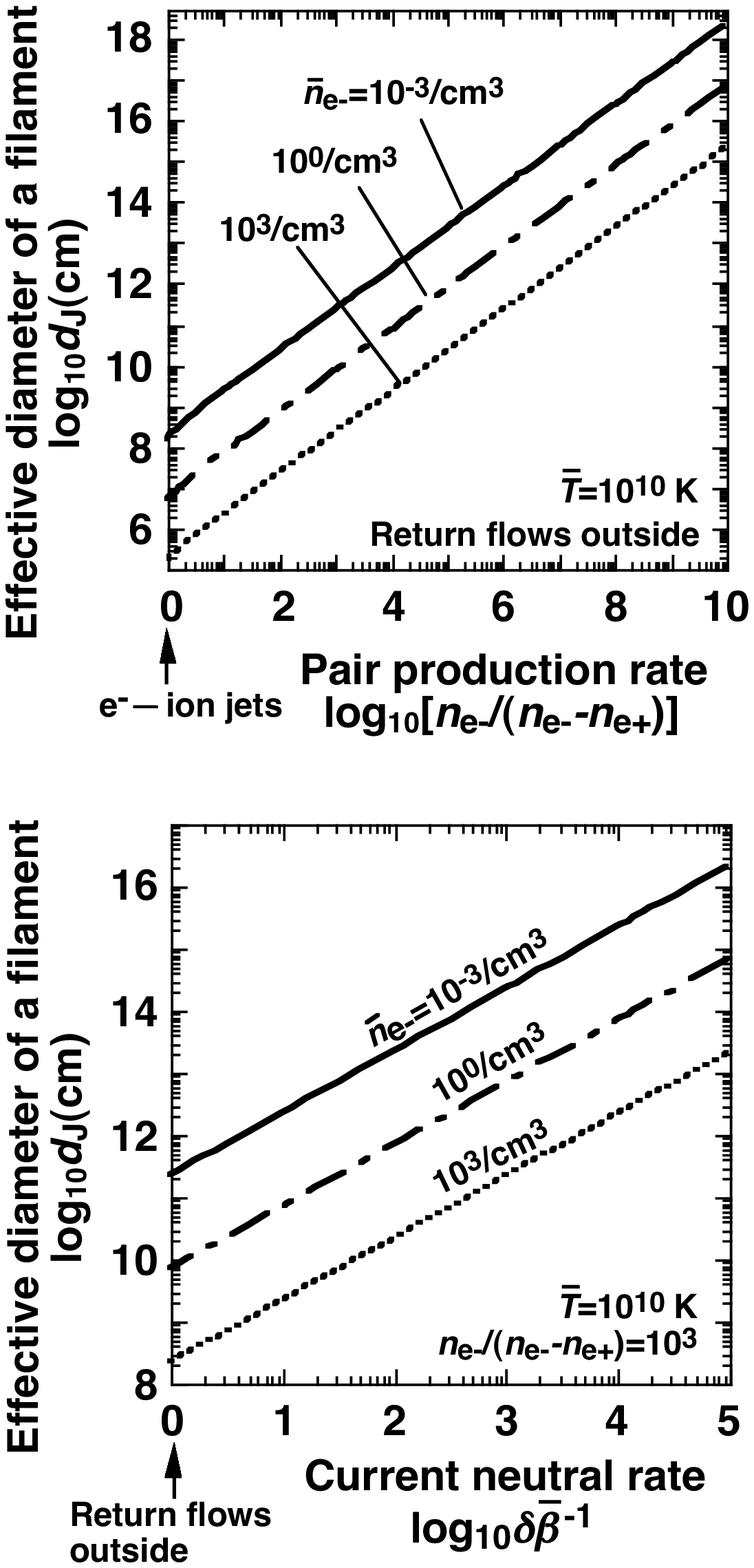}{dummy.eps}
\vspace{-1cm}
\caption{Effective diameter of a charge-neutralized
electron-positron-ion filament $d_J$ vs.
the pair production rate $\delta f_{p}^{-1}=n_{e-}/(n_{e-}-n_{e+})$ ({\it top})
and the current-neutral rate $\delta {\bar\beta}^{-1}$ ({\it bottom});
for an effective temperature of ${\bar T}=10^{10}~{\rm K}$
and average electron densities of
$\bar{n}_{e-}=10^{-3}$, $1$, and $10^3~{\rm cm^{-3}}$.
We have chosen the typical parameters of
the current-neutral rate and
the pair production rate: $\delta {\bar\beta}^{-1}=1$ ({\it top}) and
$\delta f_{p}^{-1}=10^3$ ({\it bottom}), respectively.
For further explanation see text.}
\end{figure}

\begin{figure}
\plotone{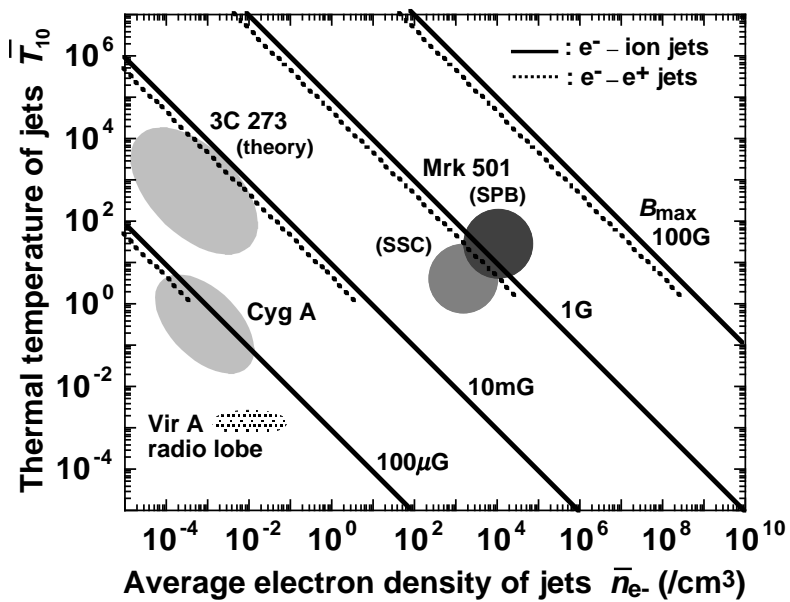}
\vspace{-2.5cm}
\caption{Maximum possible magnetic fields $B_{\theta,{\rm max}}^s$
for a given electron density $\bar{n}_{e-}$ and effective temperature
(${\bar T}_{10}\equiv {\bar T}/10^{10}~{\rm K}$) of jets.
The solid and dotted lines show the maximum fields
for the electron-ion jets ($f_p=0$)
and the electron-positron jets ($f_p=1$), respectively.
The shaded areas indicate the allowable parameter regions of
some well-known AGN jets:
Cygnus~A (Carilli et al. 1998), Virgo~A/M87 radio lobe
(Owen, Eilek, \& Kassim 2000),
3C~273 (Aharonian 2001), and Markarian~501
(predicted by the SSC model [Kataoka et al. 1999] and
by the modified SPB model [M\"ucke \& Protheroe 2001]).
Note that all quantities are in the ion rest (jet) frame.}
\end{figure}

\end{document}